\documentclass{jps-cp}
\usepackage{txfonts} 
\usepackage[svgnames]{xcolor}	
\usepackage{wrapfig}
\usepackage{multirow}
\usepackage{xcolor}
\usepackage{subfigure}
\usepackage{lineno}

\newcommand\um{{\textmu}m}

\newcommand\Rimp{$R_{\mathrm{imp}}$}
\newcommand\Ccp{$C_{\mathrm{cp}}$}

\usepackage{subfigure}
\usepackage{textcomp}
\usepackage{fouriernc}
\usepackage{here}

\title{Development of AC-LGAD detector with finer pitch electrodes for high energy physics experiments}

\author{Sayuka \textsc{Kita}$^{1}$, Koji \textsc{Nakamura}$^{2}$, Tomoka \textsc{Imamura}$^{1}$, Ikumi \textsc{Goya}$^{1}$,Kazuhiko \textsc{Hara}$^{3}$}

\inst{$^{1}$ Graduate School of Science and Technology, University of Tsukuba, 1-1-1 Tennodai, Tsukuba, Ibaraki, 305-8571, Japan \\
$^{2}$Institute of Particle and Nuclear Studies, High Energy Research Organization, 1-1 Oho, Tsukuba, Ibaraki, 305-0801, Japan \\
$^{3}$ Faculty of Pure and Applied Sciences and Tomonaga Center for the History of the Universe, University of Tsukuba, 1-1-1 Tennodai, Tsukuba, Ibaraki, 305-8571, Japan \\
}

\email{kita@hep.px.tsukuba.ac.jp, Koji.Nakamura@cern.ch}

\recdate{\today}

\abst{Low-Gain Avalanche Diode (LGAD) sensor is one of candidate sensors for the tracker at future hadron collider experiments. To use this sensor as a tracking detector, AC-LGAD sensor is being developed which has both timing and spatial resolutions. In high luminosity environments, good timing resolution (typically 30~ps) together with $\mathcal{O}$(10~\um{}) spatial resolution helps to reduce pileup effect and reconstruct tracks correctly. By optimizing fabrication parameters, 80~\um{} pitch strip and 100~\um{} pitch pixel sensors are successfully produced. 
The signal height (MPV) was 39.26$\pm$0.08~mV and 128.9$\pm$3.3~mV, respectively, for penetrating $\beta$ particles. 
The observed 60\% of signal reduction of the strip sensor is explained by a larger inter-electrode capacitance compared with the pixel sensor.
In this paper, we present the performance of fine electrode pitch AC-LGAD sensors including the pulse height and cross-talk of pixel and strip type sensors evaluated using a $\beta$-ray source and the detection efficiency measured in an 800~MeV electron beam.
}

\kword{silicon, tracker, LGAD, AC-LGAD}

\begin{document}
\maketitle

\section{Introduction}
Tracking detectors in future high-energy and high-luminosity hadron colliders are required to correctly assign the tracks associated to the hard scattering vertex among huge number of pileup vertices. Up to 200 collisions are expected in an event at the HL-LHC ATLAS experiment~\cite{atlashgtdtdr} and 10 times more in the future circular hadron collider. 
In such an environment, a revolution in the tracking system is required, e.g., realization of a time resolution of $\mathcal{O}$(10)~ps in addition to a spatial resolution of $\mathcal{O}$(10)~\um{}.

The Low-Gain Avalanche Diode (LGAD) is a semiconductor detector with improved time resolution. Presently a time resolution of 30~ps~\cite{YANG2020164379} has been achieved for minimum ionizing particles (MIPs). In realizing the spatial resolution with LGAD, we have identified potential difficulties in improving the granularity of the electrodes~\cite{HPKDCLGADRef1,HPKDCLGADRef2} without decreasing the active area fraction.  To solve the issue, a capacitively-coupled LGAD sensor (AC-LGAD)~\cite{ACLGAD2020} has been developed by KEK and University of Tsukuba in collaboration with Hamamatsu Photonics K.K. (HPK). In the previous study~\cite{ACLGAD2021}, the three process parameters, $n^+$ and $p^+$ implant concentrations and electrode coupling capacitance, were varied and the fundamental AC-LGAD performance was evaluated to optimize the parameters.

In this paper, the performance of fine electrode pitch sensors with pixel type and strip type electrodes is described. 

\section{AC-LGAD prototypes}
\begin{figure}[hbp]
    \begin{center}
        \subfigure[AC-LGAD detector]{
            \includegraphics[width=72mm]{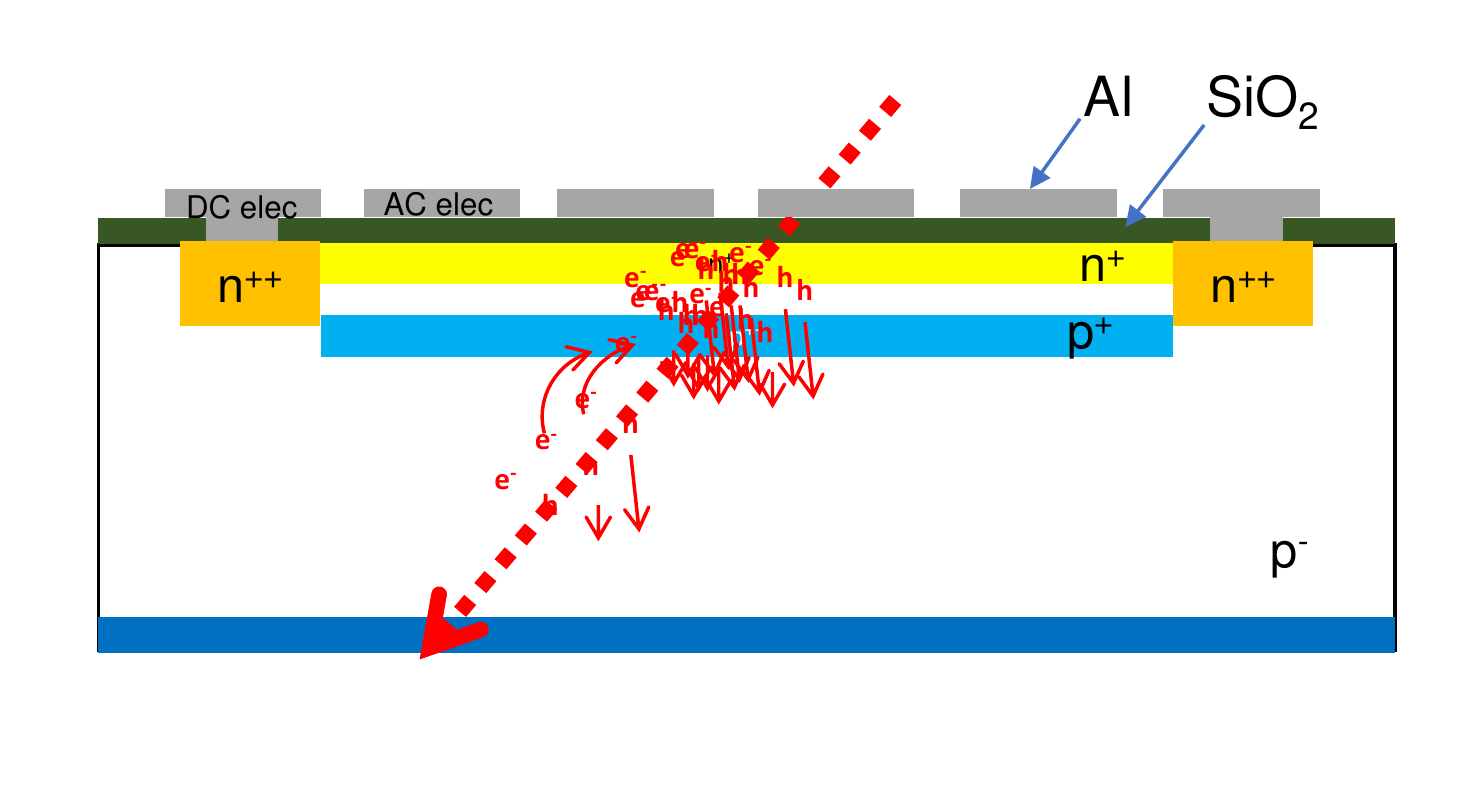}
            \label{fig:lgad}
        }
       \subfigure[Signal readout model of AC-LGAD]{
            \includegraphics[width=72mm]{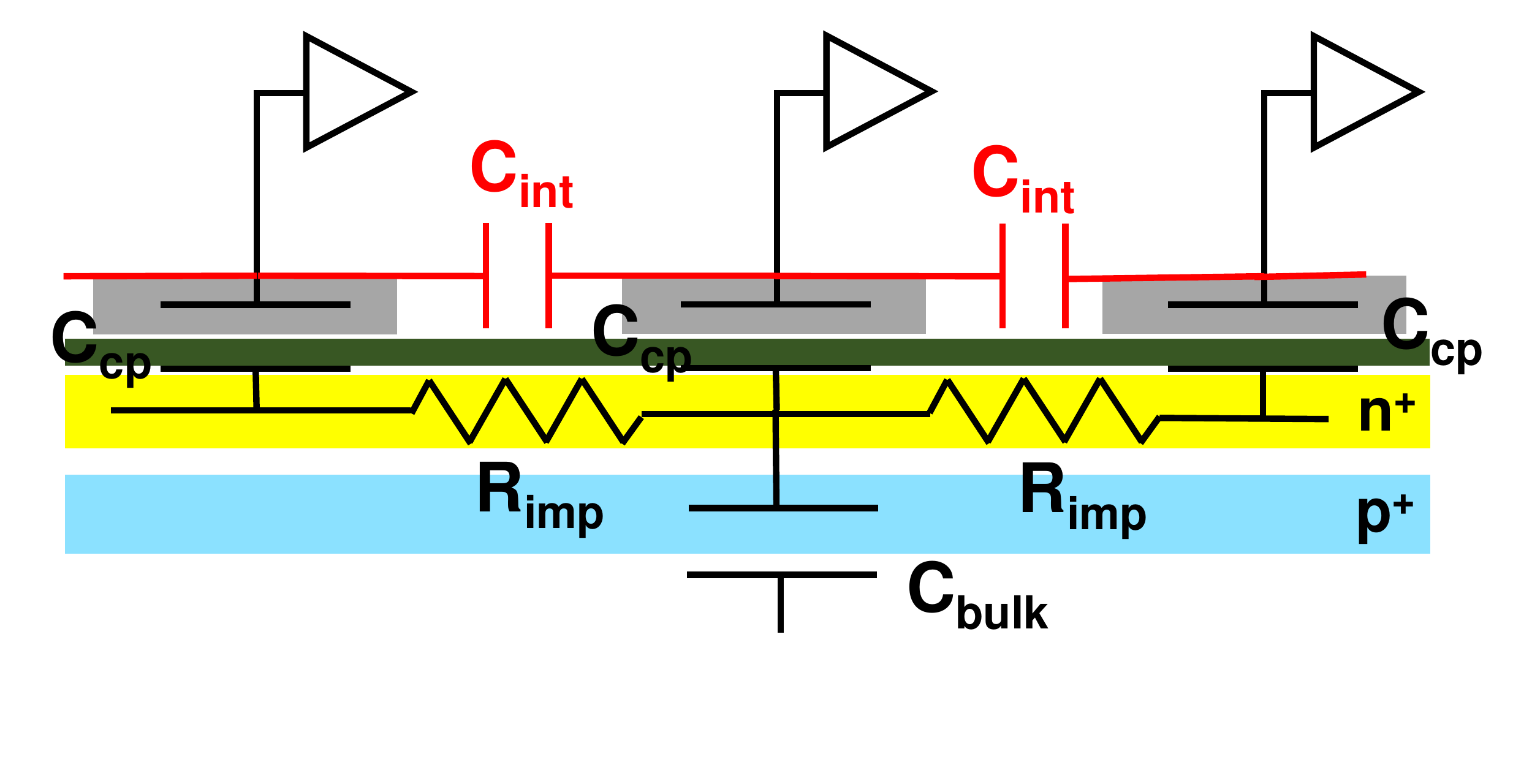}
            \label{fig:eqaclgad}
        } 
   \caption{AC-LGAD detector (left) and signal readout model of AC-LGAD (right).}
    \end{center}
\end{figure}

The LGAD sensor is basically an $n^+$-in-$p$ semiconductor diode, containing an additional $p^+$ layer under the $n^+$ electrode with a larger boron doping compared to that in the $p$-bulk region. The additional layer makes an extremely high electric field, typically over 300~kV/cm, between $n^+$ and $p^+$. Such a large electric field induces avalanche multiplication, increasing the number of electron and hole pairs by about 10-20 times of that originally produced by a MIP.  Superior time resolution is a consequence of rapid movement of a large number of electrons and holes created in the gain region by the high electric field. Concerning the spatial resolution improvement, the granularity of the electrodes is increased. In case where the gain layers (combination of $n^+$ and $p^+$) are formed separately for each direct-coupled electrode, the regions between the $n^+$ electrodes have no gain hence the fill factor is low as described in~\cite{HPKDCLGADRef2}. To overcome this feature, a capacitively-coupled LGAD (AC-LGAD) sensor, as shown in Fig.~\ref{fig:lgad},  has been developed. 
A single uniform gain layer is placed over the sensor active area and for the segmented readout, patterned aluminum electrodes are placed on the oxide layer, and the induced signal is read out capacitively. The uniform gain layer eliminates the inactive regions as the avalanche multiplication occurs uniformly across the entire active area. 

Fig.~\ref{fig:eqaclgad} illustrates the signal readout principle of the AC-LGAD sensor and equivalent circuit. 
The signal charge produced by the avalanche ($Q_{\mathrm{0}}$) flows in two paths; one is read out via coupling capacitors ($C_{\mathrm{cp}}$), and the other is cross-talk through the $n^+$ implant that has an effective resistance ($R_{\mathrm{imp}}$). The read out charge ($Q$) is determined by the impedance ratio of the two, $Z_{C_{\mathrm{cp}}}$ and $Z_{R_{\mathrm{imp}}}$, which are the impedances of $C_{\mathrm{cp}}$ and $R_{\mathrm{imp}}$, respectively. Additional cross-talk is expected via inter-electrode capacitance ($C_{\mathrm{int}}$) in case $Z_{C_{\mathrm{int}}}$ is large and becomes comparable to $Z_{C_{\mathrm{cp}}}$.
Although a presence of cross-talk to the neighboring electrodes through $n^+$ layer is a feature of AC-LGAD sensor, optimization study of $n^+$ and $p^+$ doping concentrations succeeded in suppressing the cross-talk reasonably small and keeping the signal height large, as described in~\cite{ACLGAD2020} and ~\cite{ACLGAD2021}.


According to the signal readout model, the coupling capacitance  ($C_{\mathrm{cp}}$) and inter electrode capacitance ($C_{\mathrm{int}}$) are the important parameters to be optimized while the effective $n^+$ resistance ($R_{\mathrm{imp}}$) is optimized to be as large as possible 
but avoiding the depletion to reach the insulator. The $C_{\mathrm{cp}}$ was varied by the thickness of the insulator (a combination of silicon dioxide and silicon nitride) 
layer as well as the area of the electrode. The thickness of the insulator layer was varied so that the coupling capacitance is 120~pF/mm$^2$, 240~pF/mm$^2$ and 600~pF/mm$^2$, each for the two $n^+$ resistivity points, as summarized in Table~\ref{tab:sensortype}. Since the Al electrode determines the granularity of the device, sensors with various pitches 
are tested. All the prototype sensors used for the study in this article are summarized in Table~\ref{tab:protosample}. 

\begin{table}[h]
\begin{center}
\caption{Summary of six prototype variations in $n^+$ implant concentrations and coupling capacitances, $C_{\mathrm{cp}}$.}
\label{tab:sensortype}
\begin{tabular}{c|ccc}
\hline
\hline
 & \multicolumn{3}{c}{$C_{\mathrm{cp}}$ [pF/mm$^2$]} \\
\textbf{n$^+$ resistivity} [$\Omega/\square$]  & 600  &240 & 120  \\
\hline
1600&  E600 & E240 & E120 \\
400 & C600 & C240  & C120 \\
\hline
\hline
\end{tabular}
\end{center}
\end{table}

The  $p^+$ concentration was tuned to keep the LGAD operation voltage in a range for each of the $n^+$ resistivity points.
The operation voltage is similarly 190~V among C-type senors and 160~V among E-type sensors, so the sensor performance of each type can be compared at the same bias voltage.

\begin{table}[htb]
\begin{center}
\caption{Physical parameters of the produced sensor types; Strip and Pixel. The overall size, electrode pitch, electrode width, and the numbers of electrodes in column and row directions are summarized.
"Cut Strip" contains strips of different lengths (200\um{} $\sim$ 8890\um{}) in one sensor.}
\label{tab:protosample}
\begin{tabular}{l|cc|ccc}
\hline
\hline
Sensor type & Strip & Cut Strip & \multicolumn{3}{|c}{Pixel} \\
\hline
\hline
overall dimension (mm) 
& \multicolumn{2}{c|}{11.2 $\times$ 2.7} & \multicolumn{3}{c}{2.4 $\times$ 2.4}   \\
\hline
electrode pitch (\um{})
& \multicolumn{2}{c|}{80}  &  100$\times$100 &  150$\times$150 &  200$\times$200  \\
\hline
electrode&40, 45, 60, 70  & 40, 60 &   90 &   140 &   190 \\
 dimension (\um{})   & $\times$9880 & $\times$200$\sim$8890 & $\times$90 & $\times$140 & $\times$190 \\
\hline
\#column$\times$\#row &  \multicolumn{2}{c|}{16$\times$1}   & 10$\times$10 & 6$\times$6 & 5$\times$5  \\
\hline
\hline
\end{tabular}
\end{center}
\end{table}

\section{Performance dependence on the fabrication parameters}
\subsection{Experimental procedure using $\beta$ source}
The sensor performance dependence on the sensor fabrication parameters is evaluated by injecting $^{90}$Sr $\beta$-ray as MIP. The reverse bias (negative voltage) is supplied to the rear of the sensor via conductive tape attached on a KEK 16ch amplifier board~\cite{ACLGAD2021}. The scintillation counter read out with MPPCs is used as the trigger. The amplified signals are digitized by a desktop digitizer, CAEN DT5742 (5~GHz sampling with 1024 long memory of 12-bit ADC with 1~V full scale)~\cite{ACLGAD2021}. Typical pulse shape is reported in elsewhere~~\cite{ACLGAD2020}.

\begin{figure}[htb]
    \begin{center}
        \subfigure[Pulse height distribution.]{
            \includegraphics[width=72mm]{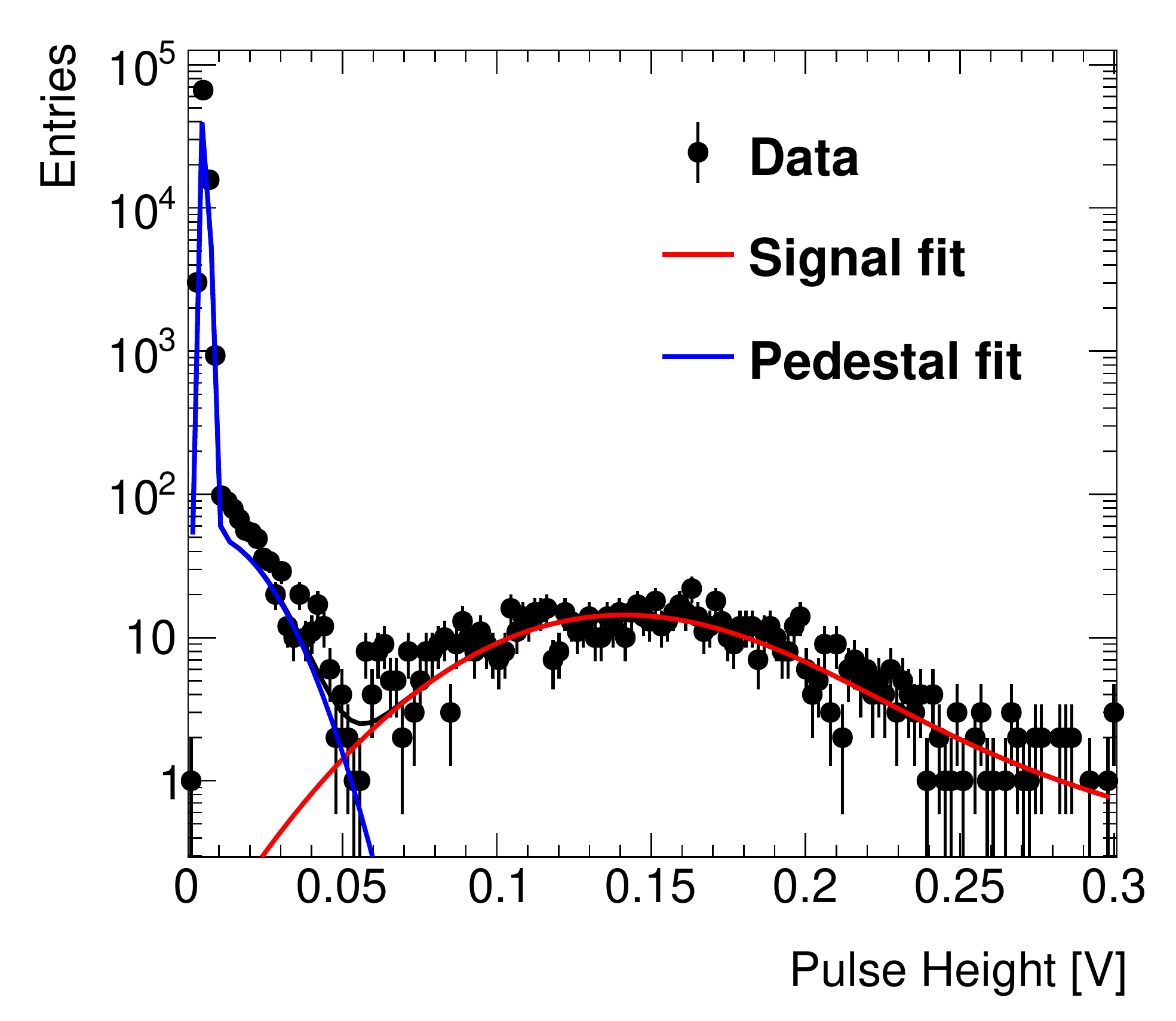}
            \label{fig:pixelph}
        }
       \subfigure[Signal efficiency and rejection factor.]{
            \includegraphics[width=65mm]{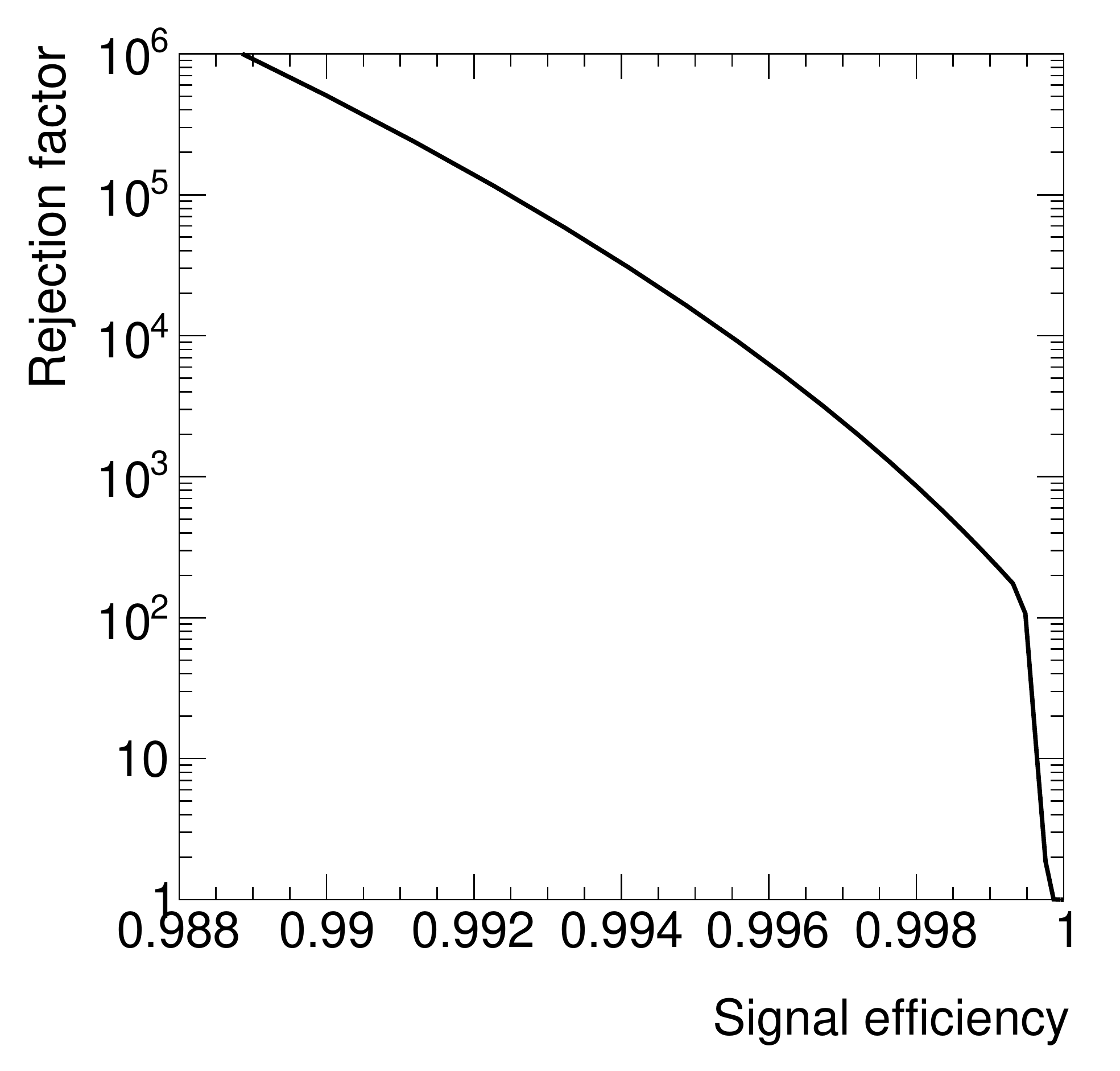}
            \label{fig:pixelsigeffi}
        } 
   \caption{(a)~Pulse height distribution of E600 pixel sensor of 100~\um{} pitch evaluated using $^{90}$Sr. Red and blue curves correspond to fitted signal and noise distributions. (b)~The signal efficiency and rejection factor of the same sensor.}
    \end{center}
\end{figure}

The signal pulse height is searched for among the sampled points as the largest negative height among the all readout channels, 4$\times$4 for the pixel and consecutive 16 for the strip sensors.
The arrival time of the signal is defined as the time at 50\% of the pulse height. 
The pulse height distributions as shown in Fig.~\ref{fig:pixelph} (and also in Fig.~\ref{fig:stripph}) are for the events whose arrival time is between 0 to 30~ns.
The signal size of the distribution is evaluated as MPV of the distribution of the signal pulse heights separated from the "noise" fitted by a Landau and Gaus convoluted function~\cite{ACLGAD2021}. The "noise" events are for the case $\beta$-ray traversed outside the readout area representing the largest noise observed in this time range. The noise distribution is fitted by an asymmetric double Gaussian function.

  Fig.~\ref{fig:pixelsigeffi} shows 
 the relationship of noise rejection factor and signal efficiency. 
 A signal efficiency of 99.54~\%  is achievable at a 10$^{-4}$ of noise rate (10$^4$ of rejection factor). 

\subsection{Pixel dependence on \Rimp{} and \Ccp{} }
\begin{figure}[htb]
    \begin{center}
        \subfigure[Dependence on \Rimp{} and \Ccp{}.]{
            \includegraphics[width=70mm]{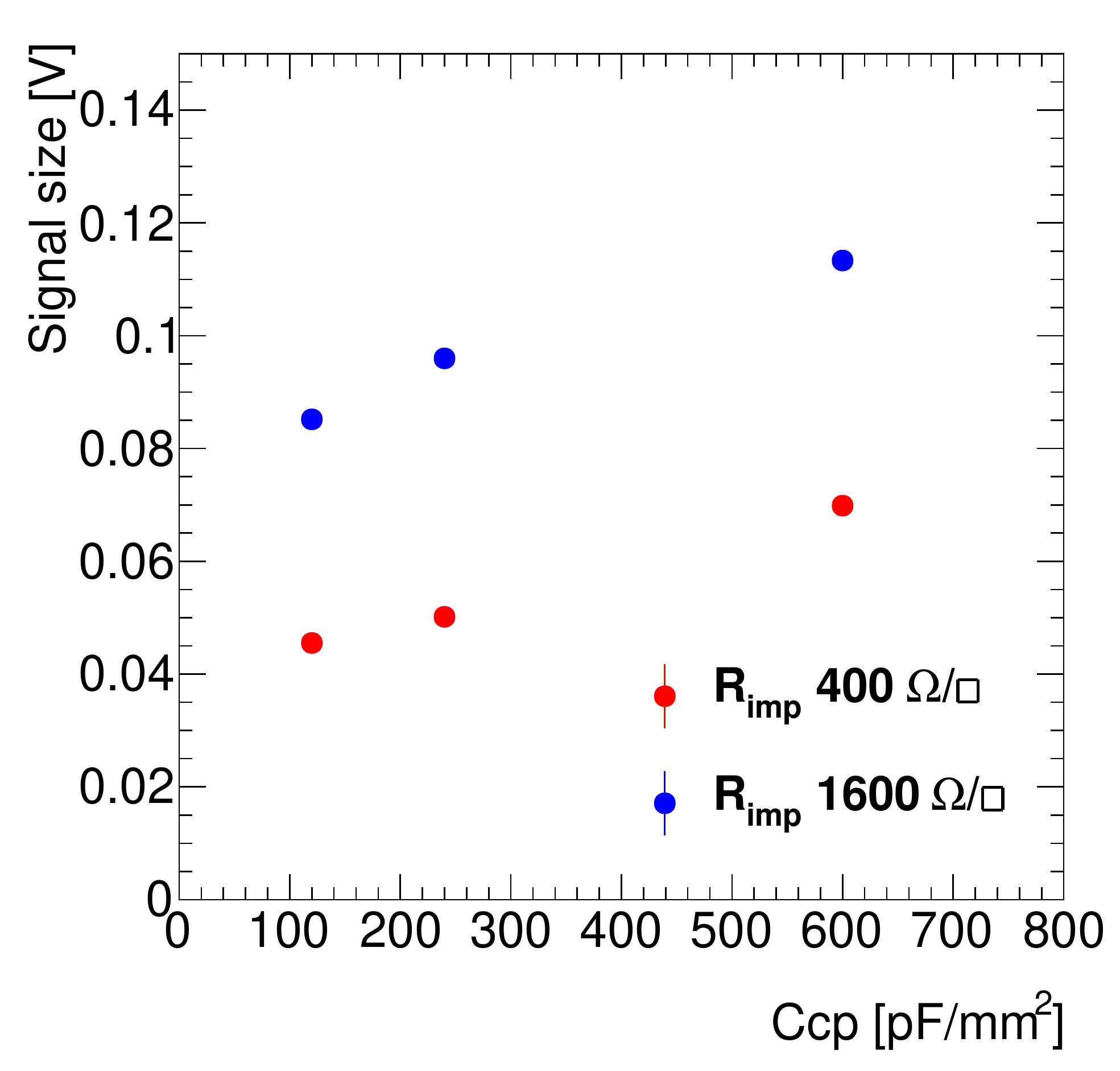}
            \label{fig:ccpcomp}
        }
       \subfigure[Dependence on pixel pitch.]{
            \includegraphics[width=70mm]{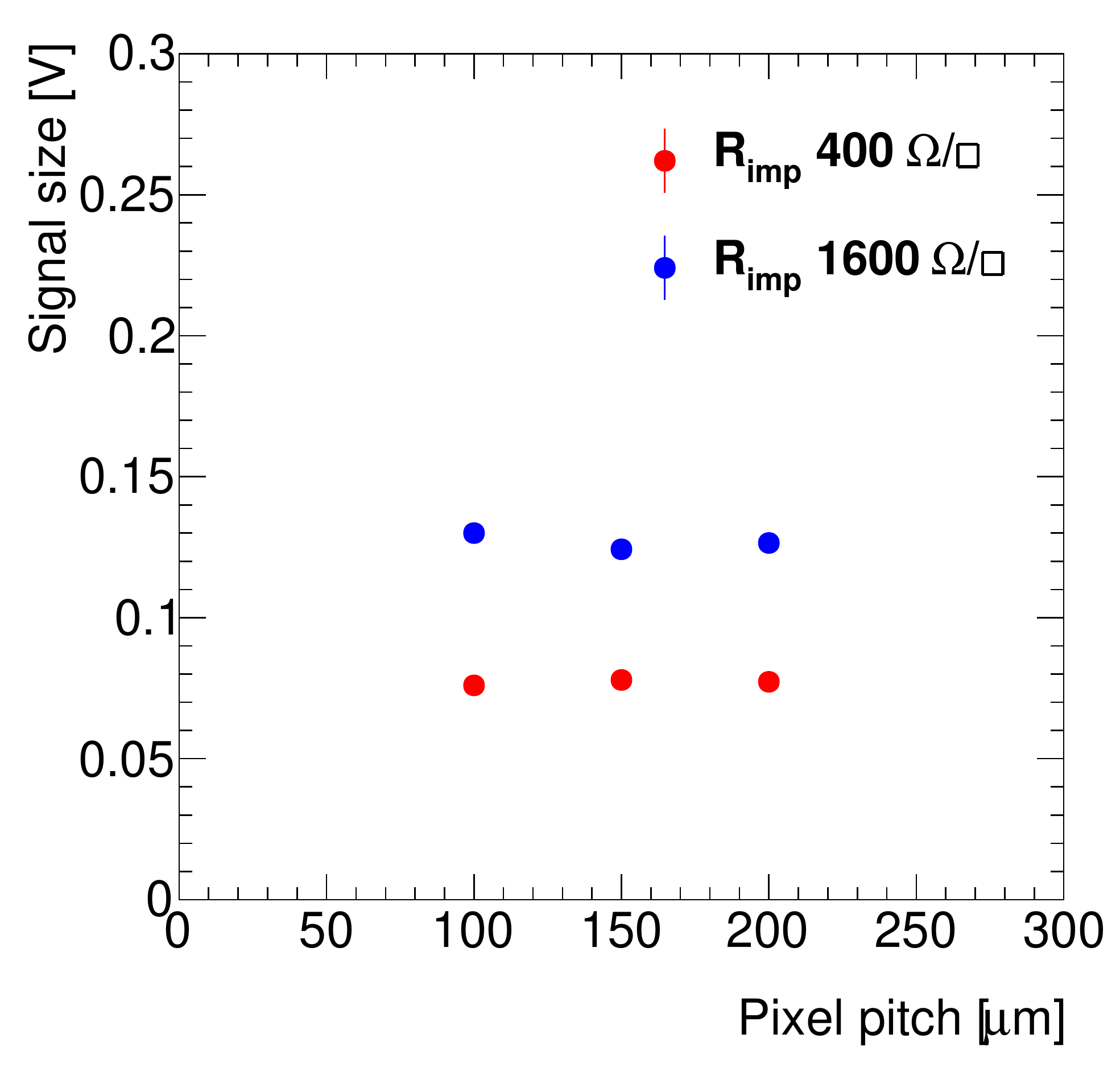}
            \label{fig:pitchcomp}
        } 
   \caption{(a)~The signal size evaluated using $^{90}$Sr as a function of \Ccp{}. Each colored point corresponds to fixed \Rimp{} value. (b)~The Signal size evaluated using $^{90}$Sr as a function of pixel pitch (\Ccp{} is 600~pF/mm$^2$). Each colored point corresponds to fixed \Rimp{} value.}
    \end{center}
\end{figure}

To measure the dependence on \Rimp{} and \Ccp{}, the six 150~\um{} pitch pixel sensors are compared for the signal size. Fig.~\ref{fig:ccpcomp} shows the signal size with respect to \Ccp{} for the two \Rimp{} values. Red and blue points correspond to \Rimp{} value of 400~$\Omega/\Box$ (C-type) and 1600~$\Omega/\Box$ (E-type), respectively. The signal size becomes larger with \Ccp{} and \Rimp{} as expected: the sensor with maximum signal size is of 1600~$\Omega/\Box$ of \Rimp{} and 600~pF/mm$^2$ of \Ccp{}. 

\subsection{Dependence on the pixel pitch}
The signal size is compared among pixel sensors with different pitches, 100, 150 and 200~\um{} and with \Ccp{} of 600~pF/mm$^2$. The signal size is observed very similar in each of the \Rimp{} values, as shown in Fig.~\ref{fig:pitchcomp}.
 
 For the  100~\um{} pixel sensor with 1600~$\Omega/\Box$ and 600~pF/mm$^2$, the obtained signal size is 128.9$\pm$3.3~mV as a fit shown in Fig.\ref{fig:pixelph}, and the separation of signal and noise distributions is excellent.

\subsection{Strip sensor performance}

\subsubsection{signal size and cross-talk}
\begin{figure}[hbp]
    \begin{center}
        \subfigure[Pulse height distribution.]{
            \includegraphics[width=70mm]{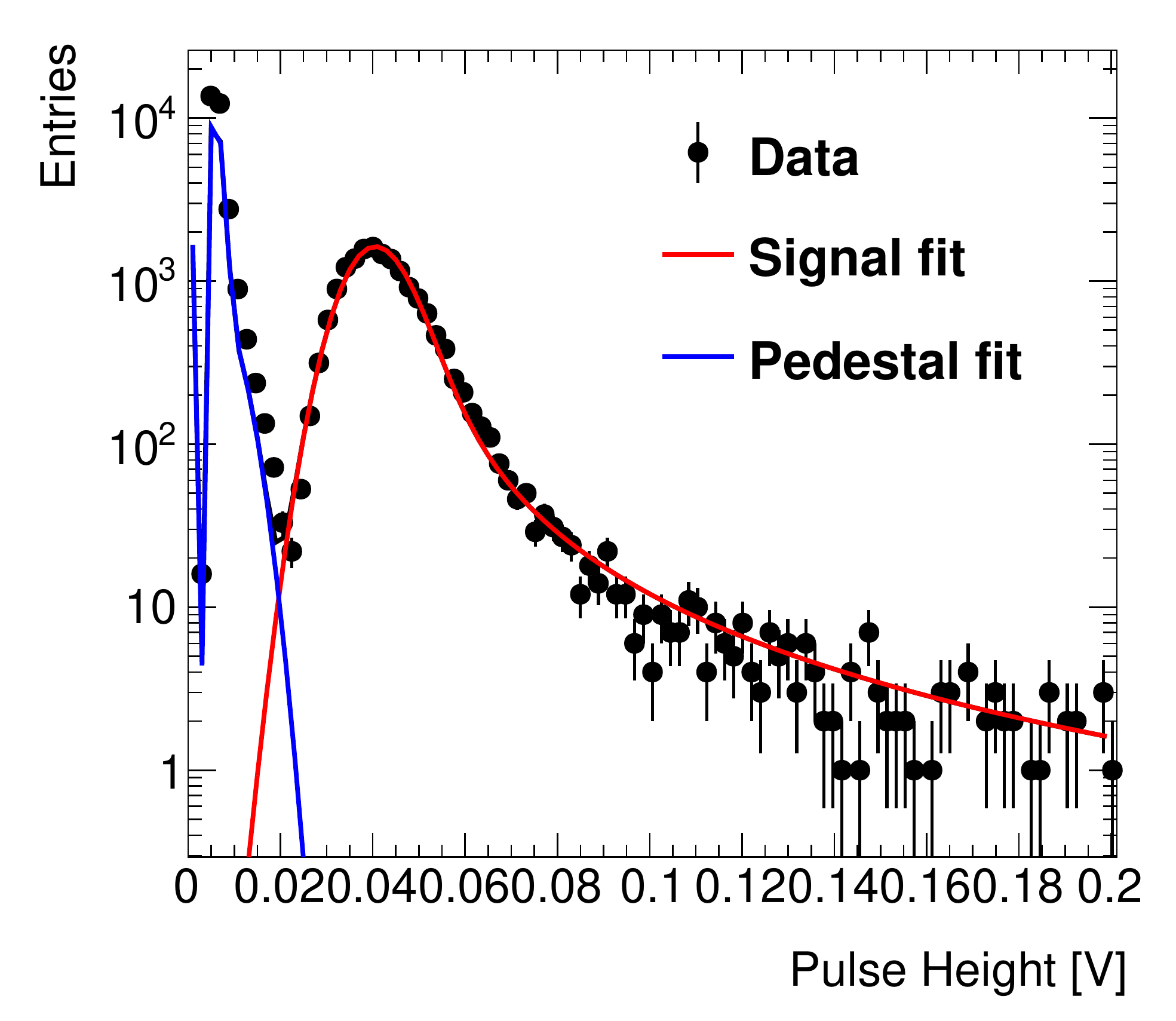}
            \label{fig:stripph}
        }
       \subfigure[Pulse height ratio to leading strip.]{
            \includegraphics[width=70mm]{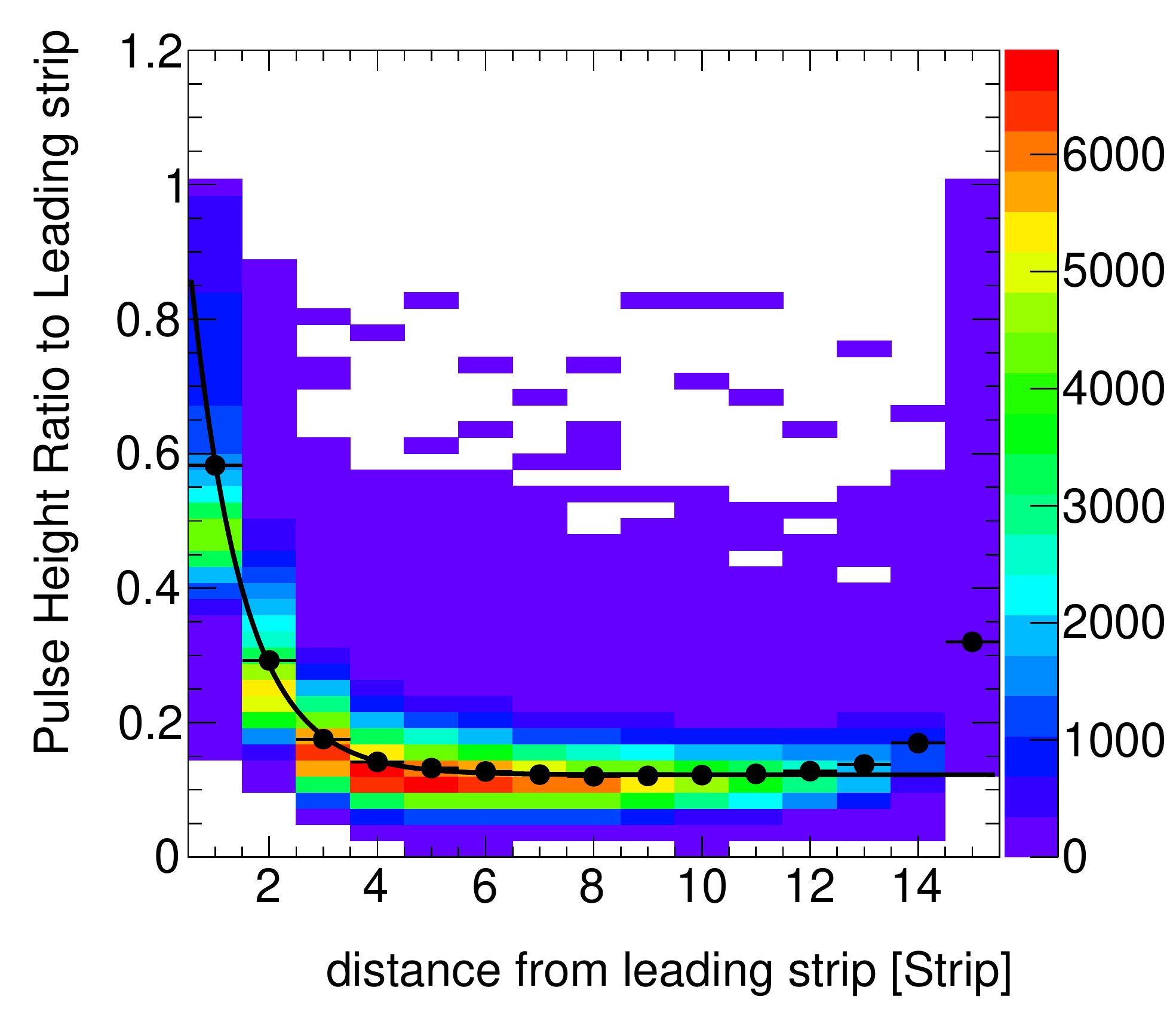}
            \label{fig:Xtalkstrip}
        } 
   \caption{(a)~Pulse height distribution of strip E600 evaluated using $^{90}$Sr. Red and blue curves are fitted signal and noise distributions. (b)~Pulse height ratios to the leading strip as function of distance from leading strip. Black dots are profiled points and fitted by an exponential + constant function.}
    \end{center}
\end{figure}

The performance of the strip sensor is also studied using $\beta$ ray as the strip sensors are useful in reducing the total number of readout channels in experiments. Fig.~\ref{fig:stripph} is the distribution of largest single channel pulse height in 16 consecutive readout strips of E600 type. The signal MPV is 39.26$\pm$0.08~mV, and signal efficiency calculated as for the pixel sensor is 99.99\% at a 10$^{-4}$ of noise rate. The magnitude of cross-talk is also evaluated as the pulse height ratio to the leading strip with the largest pulse height in an event, 
The magnitude of the cross-talk is shown in Fig.~\ref{fig:Xtalkstrip} as a function of the distance to the leading strip. The cross-talk distance is defined as the distance constant of an 
exponential function fitted to the profiled data points. The cross-talk distance of E600 strip is 62.4$\pm$0.8~\um{}, which is smaller than 1 strip distance, so the cross-talk suppression is reasonably achieved. The results indicate the designed strip sensor has a good signal to noise ratio and a small cross-talk, 
performance being expected in hadron collider experiments. Note that the non-zero constant at the tail is due to the feature of searching for the largest noise in the pre-defined time range of 30~ns.

\subsubsection{signal size dependence on the strip length}
\begin{figure}[hbp]
    \begin{center}
        \subfigure[Pulse height.]{
            \includegraphics[width=70mm]{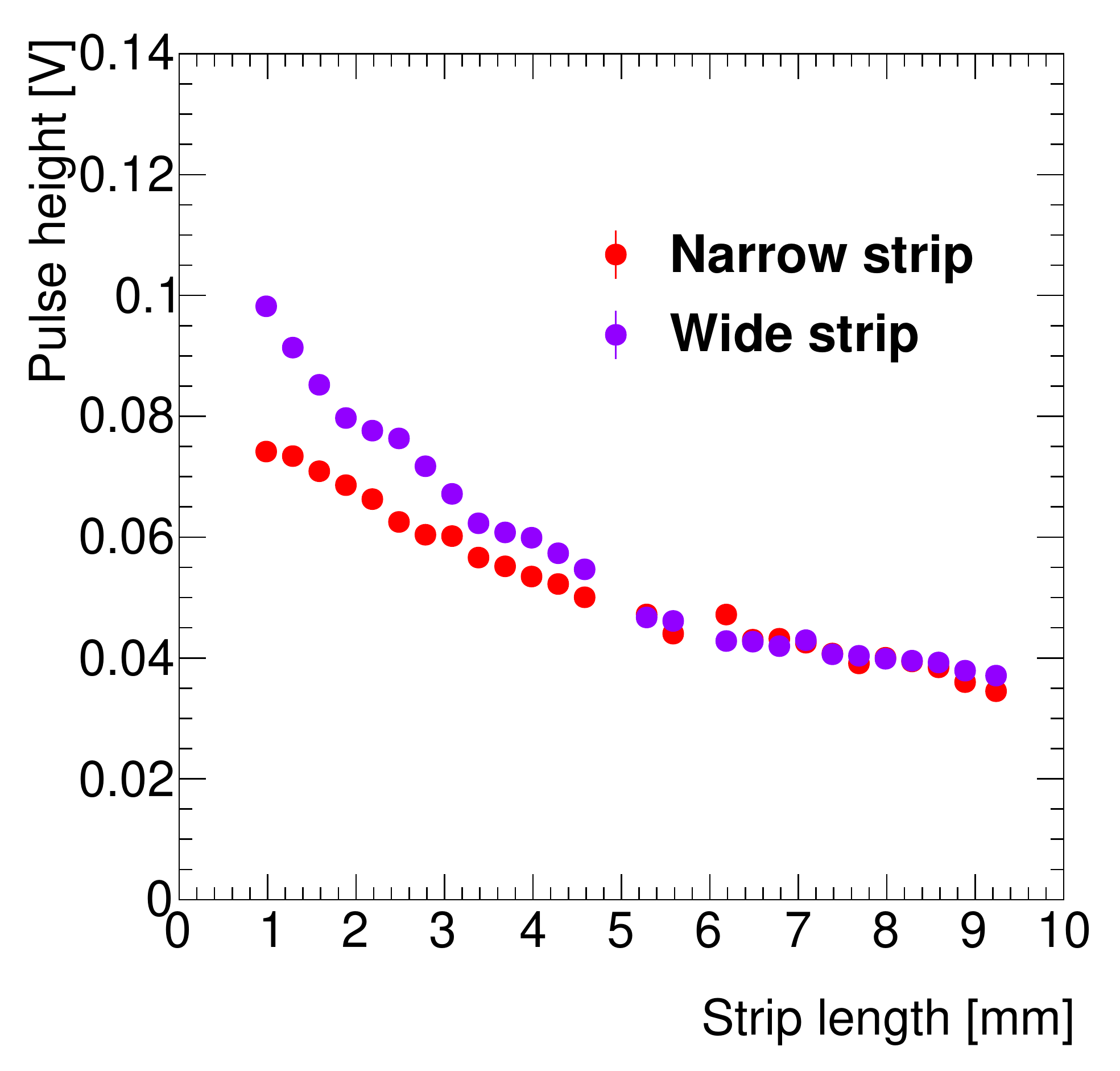}
            \label{fig:cstph}
        }
       \subfigure[Crosstalk ratio.]{
            \includegraphics[width=70mm]{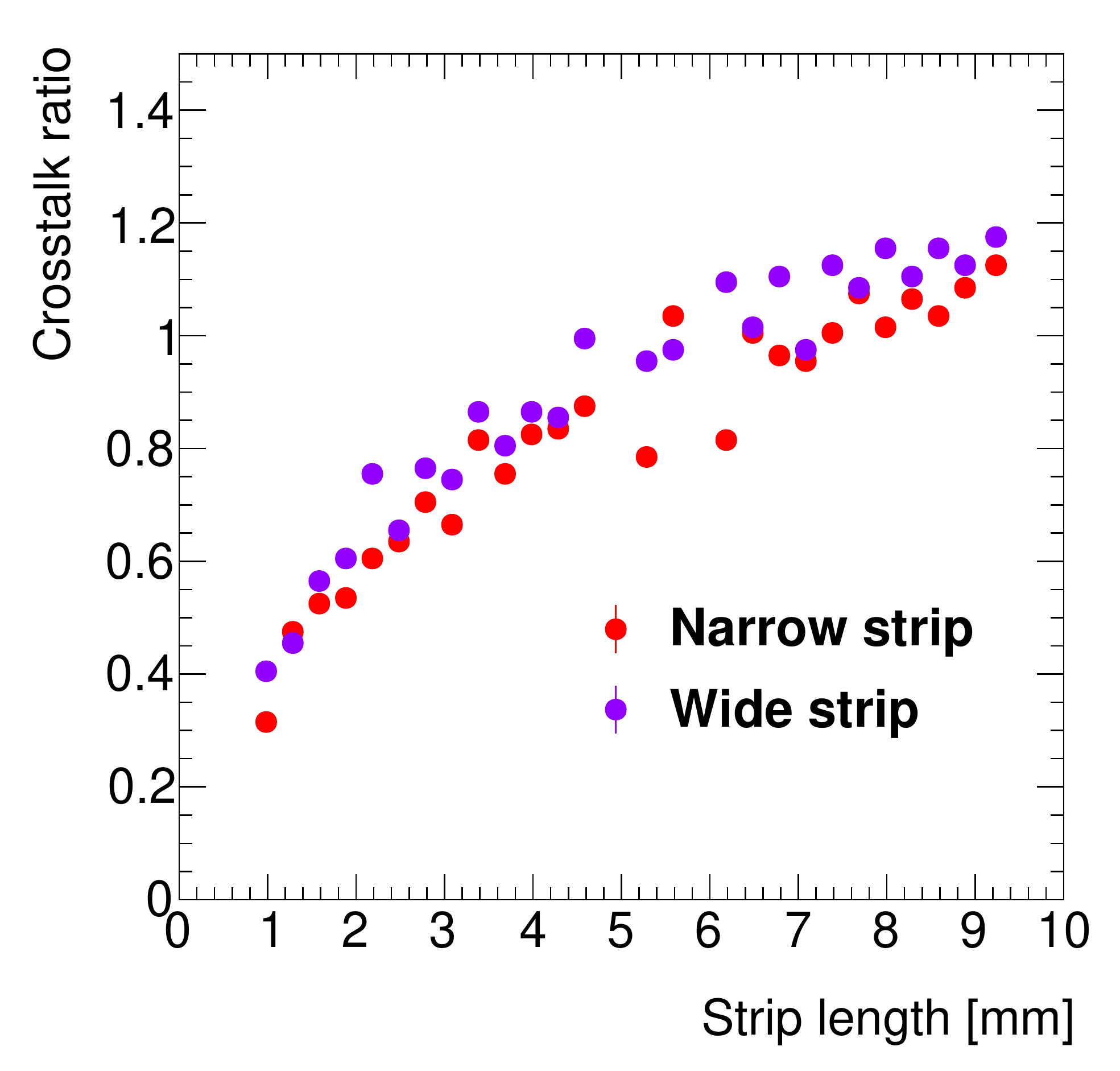}
            \label{fig:cstxt}
        } 
   \caption{The pulse height (left) and crosstalk (right) of dedicated strip sensors with different strip lengths which is measured by $^{90}$Sr. Narrow and wide strips are with electrodes of 40~\um{} and 60~\um{} widths, respectively. The cross-talk ratio is defined as pulse height of leading divided by the sum of neighboring strips.}
    \end{center}
\end{figure}

Comparing the pulse heights of the strip and pixel sensors, the strip signal size is smaller than pixel, by a factor of about 3, although the strip is larger in the electrode area, hence larger coupling capacitance. A larger inter-electrode capacitance in the strip sensor is in effect.
To check this, a dedicated strip sensor of which strip electrodes have different lengths is fabricated and response to $\beta$ ray is measured.
The strip sensor has cut-lines in diagonal direction, cutting 16 electrodes at different lengths making 32 electrodes with different lengths (as the strip is read out from both sides). In addition, we fabricated 2 electrode widths, 40~\um{} and 60~\um{}, at the same 80~\um{} pitch.
The signal size is observed to become smaller with strip length as shown in Fig.~\ref{fig:cstph}. Each colored point corresponds to the strip width of 40~\um{} (Narrow) or 60~\um{} (Wide). 
The wide strip has a larger slope, indicating the signal reduction with length is due to inter-strip capacitance, which is larger in wide electrode strips. In case of the wide strip, the signal size is reduced to 40\% from 1~mm to 10~mm strip length. 
The cross-talk is also measured. The cross-talk ratio in Fig.~\ref{fig:cstxt} is evaluated as pulse height ratio of leading strip to the sum of both neighboring strips. The cross-talk increases with strip length, and in case of wide strip, the cross-talk ratio is 1.2 at 10~mm. 
The significant increase of cross-talk contributes in reduction of signal size of the leading strip. 
The observation suggests that the signal pickup is very similar between the strip and pixel, and the strip sensor of 10~mm length is useful if the magnitude of the cross-talk is taken into account.

\section{Detection efficiency evaluated using electron beam}
\begin{figure}[hbp]
    \begin{center}
        \subfigure[Efficiency in direction X.]{
            \includegraphics[width=70mm]{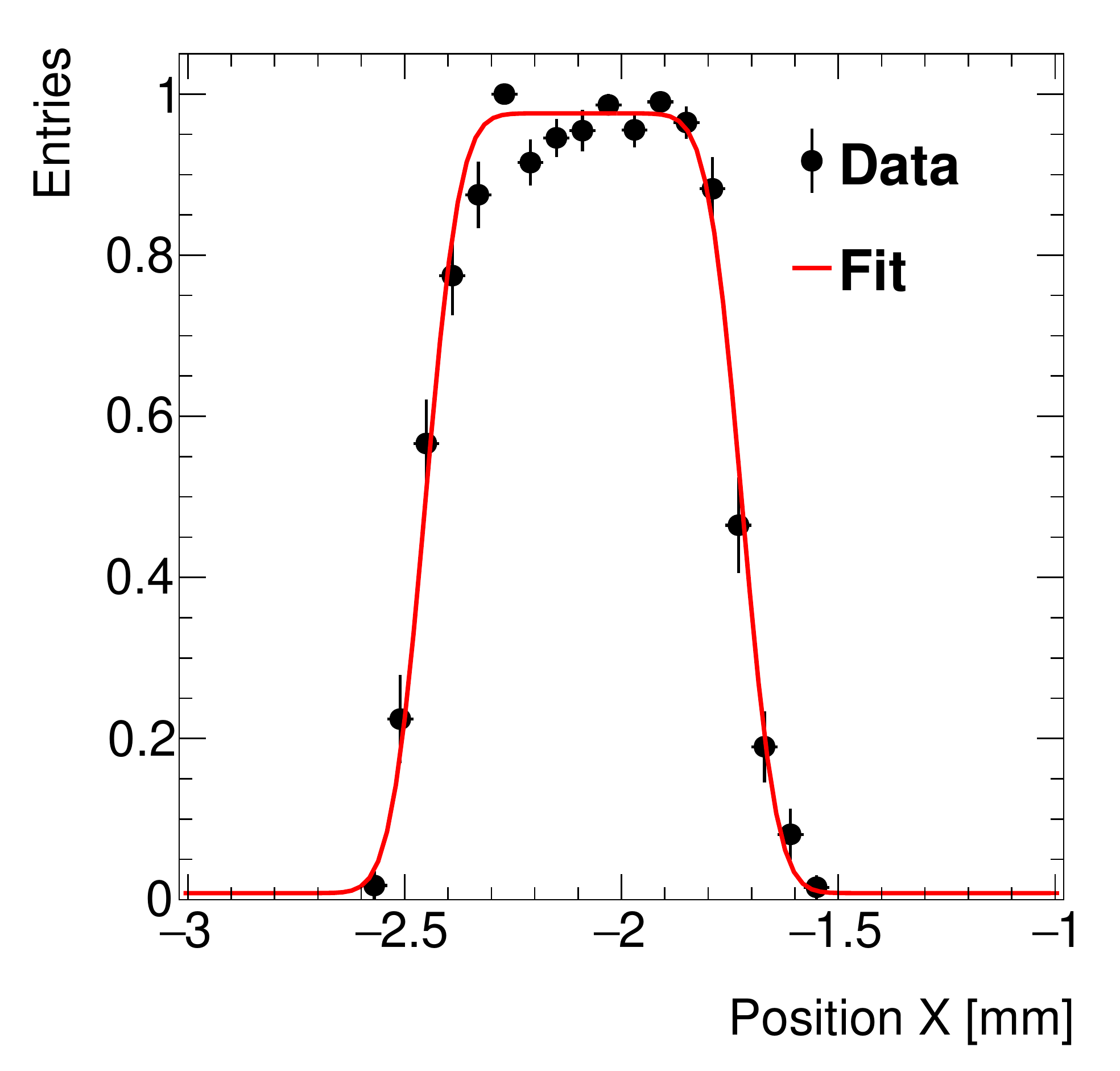}
            \label{fig:EffpixelX}
        }
       \subfigure[Efficiency in direction Y.]{
            \includegraphics[width=70mm]{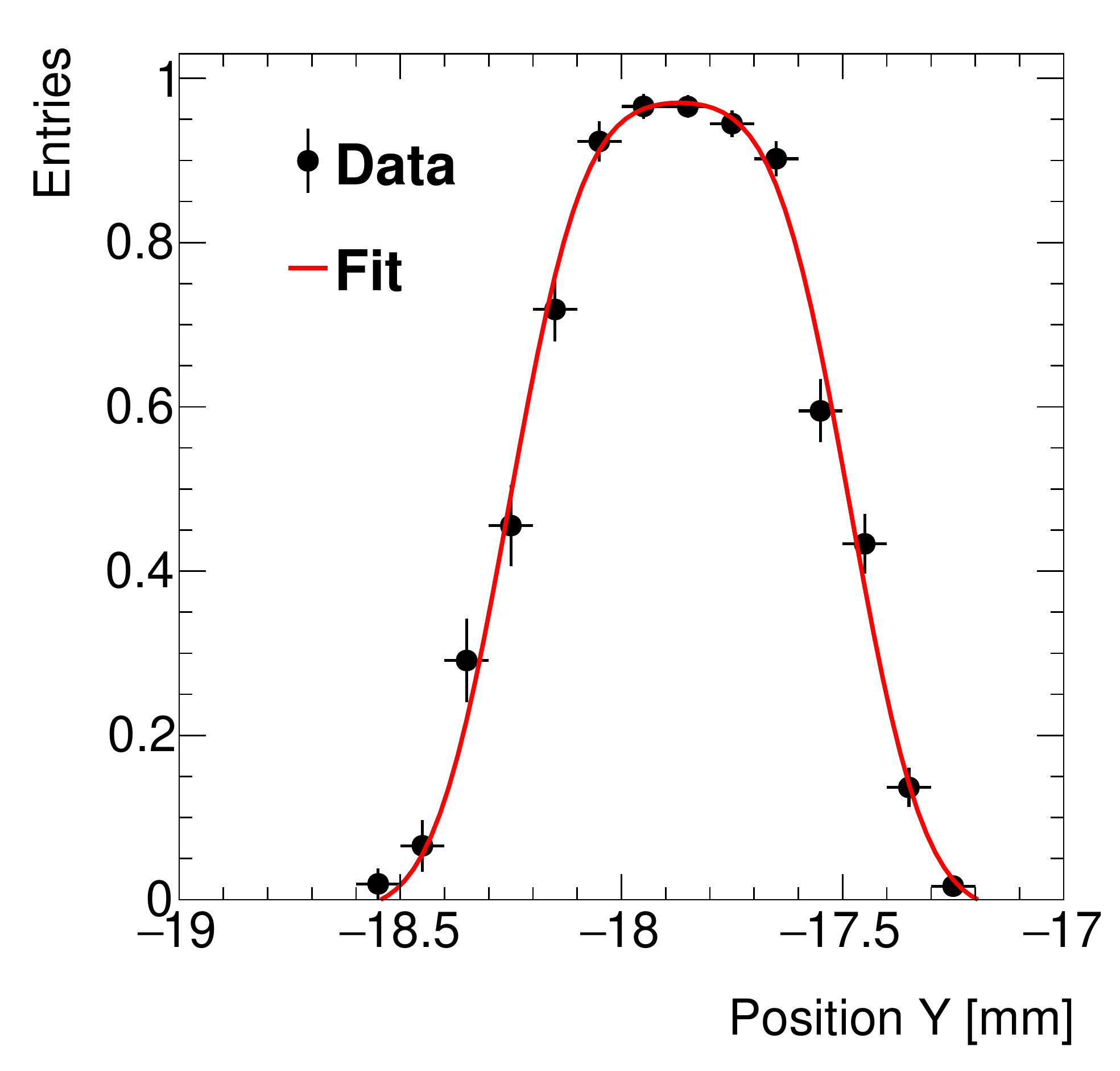}
            \label{fig:EffpixelY}
        } 
   \caption{Efficiency of 100~um pixel (E600) in direction X(left) and Y(right) evaluated in an 800~MeV electron testbeam.}
    \end{center}
\end{figure}

The efficiency of detecting MIP particles is evaluated by using an 800~MeV electron beam available at Research Center for Electron Photon Science (ELPH), Tohoku University. 
Five-layer telescope is used for tracking. The pixel size of telescope sensors is either 50$\times$250~\um{}$^2$ or 25$\times$500~\um{}$^2$. Two scintillation counters read out with MPPCs are used for trigger with one of five telescope sensors used to define region of interest (ROI) in addition. Triggering is controlled by an FPGA based trigger logic unit (TLU)~\cite{ACLGAD2021}. 
In evaluating the efficiency by 800~MeV electron testbeam, 
4$\times$4 of 100~\um{} pitch pixels located at the center of active area are wirebonded and read out with DT5742. 
The threshold for each pixel signal is determined such that the noise rate is 10$^{-4}$. 
An efficiency map is constructed for the events which recorded at least 1 hit pixel divided by the number of tracks.
The track pointing resolution is evaluated from the projected efficiency distribution across the pixel boundary edge, resulting  
48.7$\pm$7.2~\um{} in X direction and 105$\pm$12~\um{} in Y direction.
The efficiency is evaluated 
using the tracks that pass the region 1 sigma in pointing resolution inside from the pixel edges. The efficiency distributions projected in X and Y are shown in Fig.~\ref{fig:EffpixelX} and Fig.~\ref{fig:EffpixelY}. The data points are fitted by an error function on both edges, and the efficiency is evaluated in the top flat region of the distributions. The efficiency is 97.6$\pm$0.7\% in X and 97.4$\pm$1.1\% in Y, respectively. These values are high enough considering the residual multiple scattering effects.  
Note that the area with response is wider than the physical readout area of 400$\times$400~\um{}$^2$, as the surrounding pixels are not wirebonded and signals are read out via cross-talk.  

\section{Conclusions}
Development of AC-LGAD sensors with finer pitch electrodes is in progress with Hamamatsu Photonics. To make signal larger and crosstalk smaller, two parameters, $n^+$ resistivity and coupling capacitance, are optimized. The sensor which has largest $n^+$ resistivity and coupling capacitance demonstrated a best and excellent performance as a tracking sensor. 

Comparing different pixel pitches down to 100~\um{}, the pulse height is almost identical.
The signal size is 128.9$\pm$3.3~mV and signal efficiency is 99.54\% at a 10$^{-4}$ of noise rate: the signal is well separated from noise. A detection efficiency of 97\% is achieved in an 800~MeV electron beam, where the observed efficiency includes residual effect of multiple scattering.

The signal efficiency of strip is 99.99\% at a 10$^{-4}$ of noise rate. 
The inter-strip capacitance effect is larger in strip than in pixel sensor. A signal reduction to 40\% is caused at a strip length of 10~mm.

In summary, clear signal is observed 
in 100~\um{} pitch pixel and 80~\um{} pitch strip sensors, demonstrating excellent AC-LGAD sensor performance as a tracking sensor.
To deploy AC-LGAD sensors as an inner tracker in hadron collider experiments, radiation tolerance is required. New ideas of radiation hardening are investigated and will be implemented in next prototypes. Also ASIC must be developed for multiple channel readout. 

\section*{Acknowledgments}
We would like to acknowledge Hamamatsu Photonics K.K. for the fabrication of various types of AC-LGAD sensors and the discussions with the personnel have been very inspiring and fruitful. This research was partially supported by Grant-in-Aid for scientific research on advanced basic research (Grant No. 19H05193, 19H04393, 21H0073 and 21H01099) from the Ministry of Education, Culture, Sports, Science and Technology, of Japan as well as the Proposals for the U.S.-Japan Science and Technology Cooperation Program in High Energy Physics from JFY2019 to JFY2023 granted by High Energy Accelerator Research Organization (KEK) and Fermi National Accelerator Laboratory (FNAL). In conducting the present research program, the following facilities have been very important:  Research Center for ELectron PHoton Science (ELPH)
at Tohoku University.


\end{document}